\documentclass[aps,prb,twocolumn,superscriptaddress]{revtex4-2}
\usepackage{physics}
\usepackage[dvipdfmx]{graphicx}
\usepackage{hyperref}
\usepackage{here}
\usepackage{amsmath}
\usepackage{amssymb}
\usepackage{bm}
\usepackage{wrapfig}
\usepackage{ascmac}
\usepackage{mathrsfs}
\usepackage{color}
\usepackage{comment}
\usepackage[normalem]{ulem}
\usepackage{amsthm}
\usepackage{mathtools}

\theoremstyle{plain}

\theoremstyle{definition}

\theoremstyle{remark}

\theoremstyle{plain}
\newtheorem*{thm*}{Theorem}
\newtheorem*{lem*}{Lemma}
\newtheorem*{prop*}{Proposition}
\newtheorem*{cor*}{Corollary}
\newtheorem*{conj*}{Conjecture}

\theoremstyle{definition}
\newtheorem*{ass*}{Assumption}
\newtheorem*{dfn*}{Definition}

\theoremstyle{remark}
\newtheorem*{rem*}{Remark}

\newcommand{\dg}{\dagger}
\newcommand{\im}{{\rm i}}

\usepackage{comment}
\hypersetup{colorlinks=true,citecolor=blue,linkcolor=blue,urlcolor=blue}

\begin{document} 
\title{Dynamics of entanglement asymmetry for space-inversion symmetry of \\ free fermions on honeycomb lattices} 
\author{Ryogo Hara}
\email{r.hara-phys@uec.ac.jp}
\affiliation{Department of Engineering Science, University of Electro-Communications, Chofu, Tokyo 182-8585, Japan}
\author{Shimpei Endo}
\affiliation{Department of Engineering Science, University of Electro-Communications, Chofu, Tokyo 182-8585, Japan}
\affiliation{Institute for Advanced Science, University of Electro-Communications, Chofu, Tokyo 182-8585, Japan}
\author{Shion Yamashika}
\email{shion.yamashika@uec.ac.jp}
\affiliation{Department of Engineering Science, University of Electro-Communications, Chofu, Tokyo 182-8585, Japan}

\begin{abstract} 
We study the entanglement asymmetry for the space-inversion symmetry of free fermions on a two-dimensional honeycomb lattice with an on-site energy imbalance between the two sublattices. 
We show that the entanglement asymmetry of a local subsystem exhibits nonanalytic dependence on the energy imbalance, due to the presence of Dirac points in the Brillouin zone.
We also study the quench dynamics from the ground state into the inversion-symmetric point at which the energy imbalance vanishes. 
Under certain conditions on the subsystem geometry, the entanglement asymmetry relaxes to a finite value after the quench, revealing that the inversion-symmetry breaking in the initial ground state can persist even under the symmetric dynamics.
We attribute the absence of symmetry restoration to the presence of a flat energy dispersion (flat band) in a specific direction. 
\end{abstract}

\maketitle

\section{Introduction}
Nonequilibrium quantum many-body systems have attracted considerable attention for decades, as they display rich phenomena absent in equilibrium, and provide a setting to address central questions at the interface between quantum and statistical physics~\cite{polkovnikov2011colloquium,eisert2015quantum,d2016quantum,borgonovi2016quantum}.   
Although an isolated quantum system evolves unitarily and never approaches a stationary state, a local subsystem embedded in the whole system often relaxes into a statistical ensemble, which is typically a Gibbs ensemble in generic cases~\cite{deutsch1991quantum,srednicki1994chaos,srednicki1999approach,rigol2008thermalization,kaufman2016quantum,tasaki1998quantum,popescu2006entanglement,goldstein2006canonical,reimann2008foundation,linden2009quantum,short2012quantum,reimann2012equilibration,gogolin2016equilibration} or, in integrable systems, a generalized Gibbs ensemble (GGE)~\cite{rigol2007relaxation,caux2013time,ilievski2015complete,vidmar2016generalized,essler2016quench,langen2015experimental,barthel2008dephasing,cramer2008exact,calabrese2012quantum}. 
Clarifying when and how this effective equilibration occurs, and when it fails, is essential for a microscopic understanding of statistical mechanics. 
\par 
Quantum quenches provide a simple and controllable way to drive an isolated quantum system out of equilibrium. 
In this protocol, the system is initially prepared in the ground state of a given Hamiltonian and then evolves unitarily after a sudden change of its parameters. 
This setup has been widely used to study the relaxation dynamics of isolated quantum systems, including experimental realizations in trapped ion~\cite{neyenhuis2017observation,brydges2019probing,kaplan2020many} and cold atom \cite{kinoshita2006quantum,cheneau2012light,kaufman2016quantum,islam2015measuring} systems. 
A standard diagnostic in quantifying the relaxation dynamics following a quantum quench is the entanglement entropy, defined as the von Neumann entropy of a local subsystem~\cite{amico2008entanglement,eisert2010colloquium,calabrese2009entanglement,laflorencie2016quantum}. 
After a quench, it typically grows linearly in time and then approaches a stationary value equal to that of the corresponding statistical ensemble, signaling the equilibration at the subsystem level~\cite{calabrese2005evolution,fagotti2008evolution,alba2017entanglement}.
\par 
The entanglement asymmetry has been introduced as a measure of symmetry breaking within a subsystem, providing another way to quantify the relaxation dynamics from the viewpoint of symmetry~\cite{ares2023entanglement}:  
When the post-quench Hamiltonian preserves a certain symmetry while the initial state breaks it, the symmetry within a local subsystem is typically restored after the quench, as the reduced density matrix for the subsystem relaxes into a (generalized) Gibbs ensemble of the symmetric post-quench Hamiltonian. 
The entanglement asymmetry provides a quantitative means to analyze the rate of this symmetry restoration, leading to the discovery and subsequent extensive study on the quantum Mpemba effect, where a subsystem that initially breaks more the symmetry can restore it faster~\cite{ares2023entanglement,murciano2024entanglement,yamashika2024entanglement,liu2024symmetry,chalas2024multiple,caceffo2024entangled,rylands2024dynamical,joshi2024observing,rylands2024microscopic,ares2025quantum,di2025measurement,banerjee2025entanglement,banerjee2025entanglement,turkeshi2025quantum,klobas2025translation,yu2025symmetry,yu2025tuning,yamashika2025quantum,xu2025observation,yu2025review,ares2025review,teza2025speedups}. 
In addition, it has revealed counterintuitive phenomena in which the symmetry is not restored despite the symmetric dynamics~\cite{ares2023lack,yamashika2025quenching}. 
The entanglement asymmetry has also been investigated in several contexts apart from quench dynamics, including generic compact Lie groups in matrix product states~\cite{capizzi2024universal,mazzoni2025breaking}, critical systems described by conformal field theory~\cite{fossati2024entanglement,chen2024renyi,kusuki2025entanglement,fossati2025entanglement}, and Haar-random states that emulate evaporating black holes~\cite{ares2024entanglement}. 
\par 
In contrast to one-dimensional systems, the symmetry aspects of the relaxation dynamics in higher-dimensional systems remain less understood. 
In particular, previous studies have mainly focused on the simplest square-lattice systems~\cite{yamashika2024entanglement,yamashika2025quenching,yamashika2024time,gibbins2024quench,travaglino2025quasiparticle}, and the role of lattice geometry remains an open question. Moreover, while the entanglement asymmetry has been applied to physical systems with various symmetries, including those generated by non-Abelian charges~\cite{capizzi2024universal,russotto2025non}, spatial translations~\cite{klobas2025translation,gibbins2025translation}, and non-invertible symmetries~\cite{benini2025entanglement,ahmad2025many}, some fundamental symmetries in condensed matter physics, such as space-inversion and time-reversal symmetries, have been unexplored in this context.  
\par 
Free fermions on a honeycomb lattice offer an ideal platform to address these questions. 
The honeycomb lattice, consisting of two triangular sublattices, naturally admits space-inversion symmetry of exchanging the sublattices. 
This symmetry plays a crucial role in the low-energy band structure of the system:
When the inversion symmetry is preserved, the energy spectrum exhibits gapless linear dispersions at the corners of the Brillouin zone (the Dirac points), yielding massless low-energy quasiparticle excitations~\cite{castro2009electronic,wallace1947band}. On the other hand, when the symmetry is broken, for example by introducing an energy imbalance between sublattices, a gap opens and quasiparticles acquire an effective mass~\cite{semenoff1984condensed}. 
This characteristic physics, first realized in graphene~\cite{novoselov2005two,zhang2005experimental}, has stimulated extensive research into relativistic-like quasiparticles in condensed matter and emergent topological phenomena such as the quantum Hall effect~\cite{hasan2010colloquium,qi2011topological,goerbig2011electronic,castro2009electronic}. 
In addition to electronic systems, the honeycomb lattice system has been realized in ultracold atoms in optical lattices~\cite{tarruell2012creating,uehlinger2013artificial,jotzu2014experimental}, photonic systems~\cite{jacqmin2014direct,plotnik2014observation,lu2014topological,ozawa2019topological}, and mechanical metamaterials~\cite{kariyado2015manipulation,nash2015topological}. 
\par 
In this paper, we study the quench dynamics of the entanglement asymmetry associated with the space-inversion symmetry in spinless (i.e., spin-polarized) free fermions on a honeycomb lattice. 
We derive analytical expressions for the time evolution of the entanglement asymmetry in the quench dynamics starting from the ground state of the Hamiltonian with the on-site energy imbalance between the sublattices into the inversion-symmetric Hamiltonian without the imbalance. 
When size of the subsystem, taken as a periodic stripe shape, is odd in the periodic direction, the entanglement asymmetry tends to zero and the inversion symmetry is restored after the quench, as the subsystem relaxes into the GGE of the inversion-symmetric post-quench Hamiltonian. 
In contrast, when the subsystem size is even, the entanglement asymmetry relaxes to a finite value, indicating the absence of inversion-symmetry restoration. 
We attribute this absence of symmetry restoration to a macroscopic occupation of quasiparticle modes with zero group velocity (i.e., flat band), demonstrating the crucial role of band structures in the relaxation dynamics.
\par 
This paper is organized as follows. In Sec.~\ref{sec:setup}, we introduce our quench protocol and define the entanglement asymmetry for the space-inversion symmetry. In Sec.~\ref{sec:dimensional reduction}, we describe the method to calculate the entanglement asymmetry in the two-dimensional system. In Sec.~\ref{sec:result_GS}, we analyze the entanglement asymmetry in the ground state of the Hamiltonian with the on-site energy imbalance between sublattices. 
In Sec.~\ref{sec:result_TEV}, we examine its time evolution after the quench and show that, in certain cases, the inversion symmetry is not restored even in the large time limit. 
In Sec.~\ref{sec:lack}, we identify the physical mechanism responsible for the absence of inversion-symmetry restoration. 
Section~\ref{sec:conclusion} summarizes our results. 
Appendixes provide technical details and derivations of several formulas used in the main text. 
We set $\hbar=1$ and lattice constant unity throughout this paper. 

\section{System setup and Definitions of entanglement asymmetry}\label{sec:setup}
\begin{figure}
\includegraphics[width=0.9\linewidth]{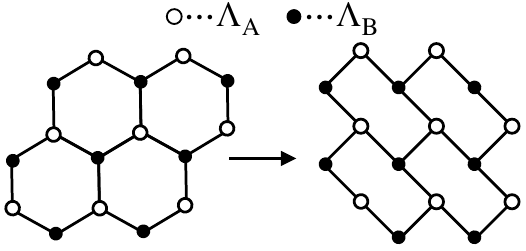}
\caption{Schematic illustration of the transformation from the honeycomb to the brickwork lattices.}\label{fig:lattice}
\end{figure}

We consider free fermions on a honeycomb lattice consisting of triangular sublattices $\Lambda_{\rm A}$ and $\Lambda_{\rm B}$. 
For convenience, we perform the lattice transformation from the honeycomb lattice to a brickwork lattice as shown in Fig.~\ref{fig:lattice}, which does not change the lattice topology. 
The system is described by the following Hamiltonian 
\begin{multline}
    H_0=-J\sum_{\mathbf{i} \in \Lambda_{\rm A}} 
    \sum_{\nu=1,2,3} 
    (a_\mathbf{i}^\dag  b_{\mathbf{i}+\mathbf{d}_\nu} +\mathrm{H.c.})
    \\
    +
    M \sum_{\mathbf{i}\in \Lambda_{\rm A}} 
    a_\mathbf{i}^\dag a_\mathbf{i}
    -M\sum_{\mathbf{i}\in \Lambda_{\rm B}} b_{\mathbf{i}}^\dag b_{\mathbf{i}},
    \label{eq:H_0}
\end{multline}
where $a_\mathbf{i}~(b_\mathbf{i})$ is the annihilation operator of a fermion on $\mathbf{i}\in \mathrm{\Lambda}_{\rm A(B)}$, $\mathbf{i}=(i_x,i_y)$ is the two-dimensional vector identifying the position of the site, $J>0$ is the hopping amplitude between nearest-neighbor sites, $M$ is the on-site energy difference between the two sublattices, and $\mathbf{d}_\nu~(\nu=1,2,3)$ are the nearest-neighbor vectors that are defined as (we take the lattice constant as a unit of length)
\begin{align}
    \mathbf{d}_1=\frac{1}{2}\mqty(1\\-1),
    \mathbf{d}_2=\frac{1}{2}\mqty(-1\\1),
    \mathbf{d}_3=\frac{1}{2}\mqty(-1\\-1).
\end{align}
We denote as $L_{x(y)}$ the length of the system in the longitudinal (transverse) direction. The periodic boundary conditions are imposed along both directions. 
When $M/J=0$, the Hamiltonian~\eqref{eq:H_0} is invariant under the space-inversion $\mathcal{P}$ that exchanges sublattices $\Lambda_{\rm A}$ and $\Lambda_{\rm B}$  as $\mathcal{P}a_\mathbf{i}\mathcal{P}^{-1}=b_{-\mathbf{i}}$ and $\mathcal{P}b_\mathbf{i}\mathcal{P}^{-1}=a_{-\mathbf{i}}$, whereas this inversion symmetry is broken for $M/J\neq0$.  
The inversion-symmetry breaking changes the energy spectrum of the Hamiltonian~\eqref{eq:H_0} from gapless to gapped, as we will see below. 
\begin{figure*}
    \includegraphics[width=0.99\linewidth]{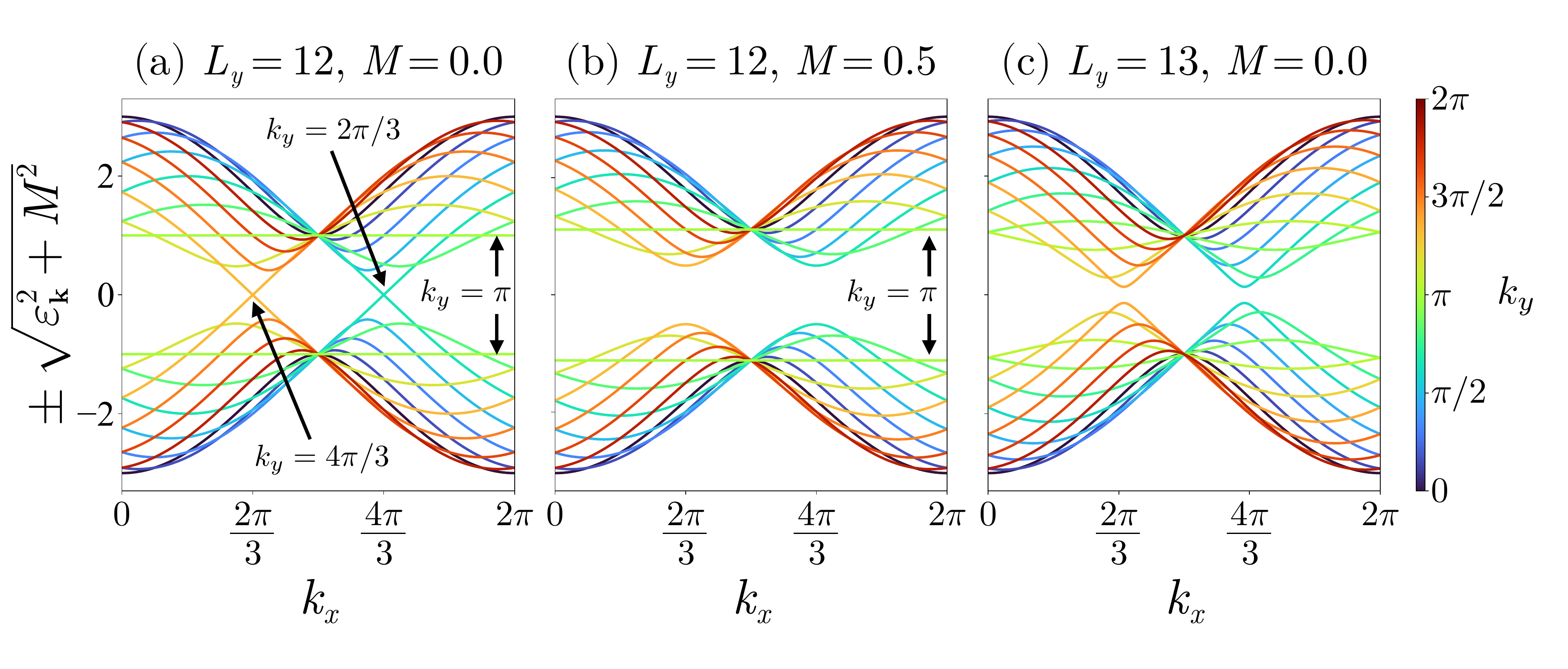}
    \caption{The energy spectra of the Hamiltonian in Eq.~\eqref{eq:H_0} for several fixed values of $L_y$. We set $J=1$ in all the plots.}
    \label{fig:band}
\end{figure*}
\par 
Performing the Fourier transformations
\begin{align}
    a_\mathbf{i}
    &= 
    \frac{1}{\sqrt{L_xL_y}} \sum_\mathbf{k} e^{\im \mathbf{k}\cdot \mathbf{i}} a_\mathbf{k}, \\
    b_\mathbf{i}
    &= 
    \frac{1}{\sqrt{L_xL_y}} \sum_\mathbf{k} e^{\im \mathbf{k}\cdot \mathbf{i}} b_\mathbf{k},
\end{align}
where $\mathbf{k}=(k_x,k_y)$ with $k_{x(y)}= 2\pi n_{x(y)} /L_{x(y)}~(n_{x(y)}=0,...,L_{x(y)}-1)$, the Hamiltonian~\eqref{eq:H_0} reduces to 
\begin{align}
    H_0
    = 
    -J
    \sum_\mathbf{k}
    (a_\mathbf{k}^\dag,b_\mathbf{k}^\dag)
    \mqty[-M/J & \sum \limits_{\nu=1}^3 e^{\im \mathbf{k}\cdot \mathbf{d}_\nu} 
    \\
    \sum \limits_{\nu=1}^3 e^{-\im \mathbf{k}\cdot \mathbf{d}_\nu} 
    & 
    M/J
    ]
    \mqty(a_\mathbf{k} \\ b_\mathbf{k}).
    \label{eq:H_0k}
\end{align}
It can be diagonalized by the linear transformation 
\begin{align}
    \mqty(\alpha_{\mathbf{k}} \\ \beta_{\mathbf{k}})
    = 
    \mqty(
    e^{\im \frac{\phi_\mathbf{k}}{2}}\cos \frac{\theta_\mathbf{k}}{2} & 
    e^{-\im \frac{\phi_\mathbf{k}}{2}}\sin \frac{\theta_\mathbf{k}}{2}
    \\
    -e^{\im \frac{\phi_\mathbf{k}}{2}}\sin \frac{\theta_\mathbf{k}}{2} & 
    e^{-\im \frac{\phi_\mathbf{k}}{2}}\cos \frac{\theta_\mathbf{k}}{2}   
    )\mqty(a_\mathbf{k}\\ b_\mathbf{k}),
    \label{eq:bogoliubov}
\end{align}
where the angles $(\theta_\mathbf{k},\phi_\mathbf{k})$ are determined by
\begin{align}
    \cos\theta_\mathbf{k}&=\frac{M}{\sqrt{\varepsilon_\mathbf{k}^2+M^2}},
    \label{eq:angles}   
    \\
    \sin\theta_\mathbf{k}\cos\phi_\mathbf{k}
    &=-\frac{J\sum_{\nu=1}^3\cos(\mathbf{k\cdot d}_\nu)}{\sqrt{\varepsilon_\mathbf{k}^2+M^2}}, 
    \\
    \sin\theta_\mathbf{k}\sin\phi_\mathbf{k}
    &=\frac{J\sum_{\nu=1}^3\sin(\mathbf{k\cdot d}_\nu)}{\sqrt{\varepsilon_\mathbf{k}^2+M^2}},
\end{align}
with 
\begin{align}
    \varepsilon_\mathbf{k}=J\abs{\sum_{\nu=1}^3 e^{\im \mathbf{k\cdot d_\nu}}}. 
\end{align}
Substituting Eq.~\eqref{eq:bogoliubov} into Eq.~\eqref{eq:H_0k}, we obtain 
\begin{align}
    H_0 = \sum_\mathbf{k} \sqrt{\varepsilon_\mathbf{k}^2+M^2}(\alpha_\mathbf{k}^\dag \alpha_\mathbf{k}-\beta_\mathbf{k}^\dag \beta_\mathbf{k}).
    \label{eq:H_0_diagonal}
\end{align}
\par 
The above expression shows that the energy spectrum of the Hamiltonian consists of upper and lower bands with dispersions $+\sqrt{\varepsilon_\mathbf{k}^2+M^2}$ and $-\sqrt{\varepsilon_\mathbf{k}^2+M^2}$, respectively, and that
$\alpha_\mathbf{k}$ and $\beta_\mathbf{k}$ correspond to the quasiparticle operators associated with these bands.
If we denote as $\ket{\Psi_0}$ the ground state of the Hamiltonian~\eqref{eq:H_0_diagonal} at half filling, it corresponds to the fully occupied lower band and the empty upper band, i.e., 
\begin{align}
\ket{\Psi_0} = \prod_\mathbf{ k} \beta_\mathbf{k}^\dag \ket{0}, \label{eq:Psi_0}
\end{align}
where $\ket{0}$ is the fermionic vacuum state annihilated by $a_\mathbf{i}$ and $b_\mathbf{i}$, i.e., $a_\mathbf{i}\ket{0}=b_\mathbf{i}\ket{0}=0~\forall\mathbf{i}$. 
\par 
Figure~\ref{fig:band} shows the energy spectra of the Hamiltonian~\eqref{eq:H_0_diagonal}. 
As shown in Fig.~\ref{fig:band} (a), when $M=0$ and $L_y$ is a multiple of three, linear dispersions appear around the Dirac points $\mathbf{k}=\pm (2\pi/3,4\pi/3)$~\cite{castro2009electronic}, where $-k_{x(y)}$ and $2\pi -k_{x(y)}$ are equivalent because of the periodicity of the Brillouin zone. 
As seen in Fig.~\ref{fig:band}~(b), a gap opens when $M\neq0$ as a consequence of inversion-symmetry breaking~\cite{semenoff1984condensed}. 
Even when $M=0$, a gap opens if $L_y$ is not divisible by three because the quantized momenta $\mathbf{k}$ do not take the Dirac points, as shown in Fig.~\ref{fig:band}~(c). 
We also find in Figs.~\ref{fig:band}~(a) and (b) that $\varepsilon_\mathbf{k}$ with fixed $k_y=\pi$ becomes independent of $k_x$, which occurs only when $L_y$ is even.  
The presence of the flat dispersion relation in the $k_x$-direction critically affects the relaxation dynamics after a quantum quench, as shown in Secs.~\ref{sec:result_TEV} and \ref{sec:lack}.
\par
\begin{figure}
    \centering
    \includegraphics[width=0.95\linewidth]{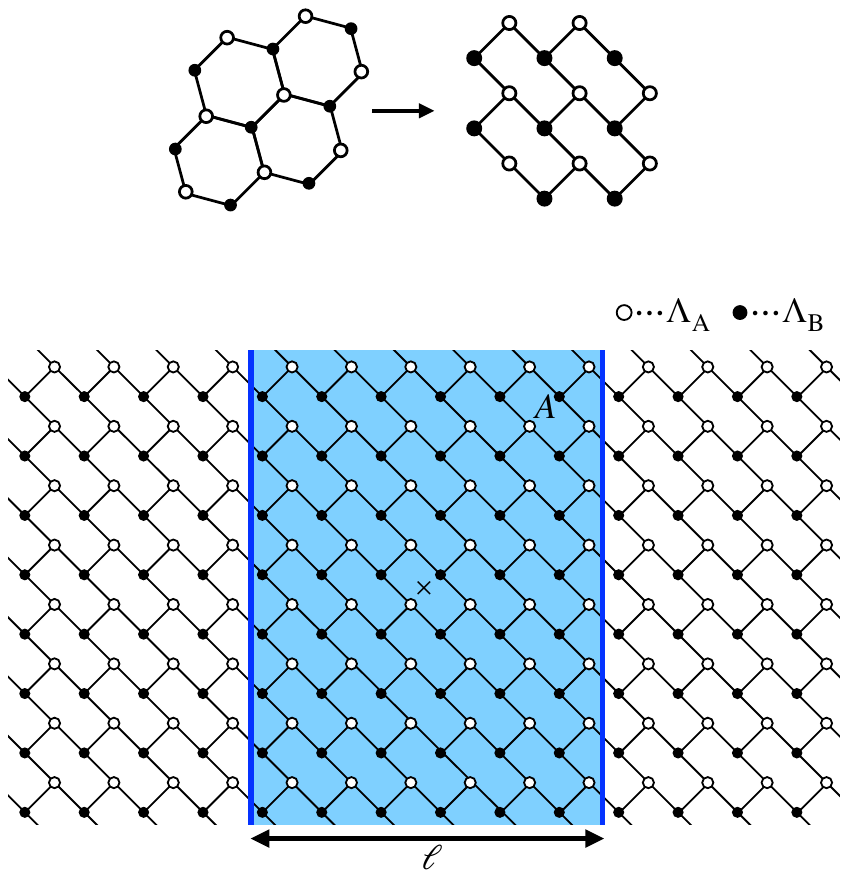}
    \caption{Schematic illustration of the bipartition of the whole system into the subsystems. 
    The blue region represents subsystem $A$ and the cross symbol at the center indicates the coordinate origin. }
    \label{fig:subsystem}
\end{figure}

 
We study the dynamical behavior of inversion-symmetry breaking following the global quantum quench from an asymmetric point $M\neq0$ into the symmetric point $M=0$. 
After the quench, the system is described by the time-evolved state $\ket{\Psi_t}=e^{-\im t H}\ket{\Psi_0}$, where $\ket{\Psi_0}$ and $H$ are the initial ground state given in Eq.~\eqref{eq:Psi_0} and the post-quench Hamiltonian that is obtained by setting $M=0$ into Eq.~\eqref{eq:H_0}, respectively. 
The post-quench Hamiltonian specifically reads 
\begin{align}
    H=\sum_\mathbf{k}\varepsilon_\mathbf{k}
    (\gamma_{\mathbf{k}+}^\dag
    \gamma_{\mathbf{k}+}
    -
    \gamma_{\mathbf{k}-}^\dag
    \gamma_{\mathbf{k}-}),
    \label{eq:H}
\end{align}
with 
\begin{align}
\mqty(\gamma_{\mathbf{k}+} \\ \gamma_{\mathbf{k}-})
    = 
    \frac{1}{\sqrt{2}}
    \mqty(
    e^{\im \frac{\phi_\mathbf{k}}{2}} & 
    e^{-\im \frac{\phi_\mathbf{k}}{2}}
    \\
    -e^{\im \frac{\phi_\mathbf{k}}{2}}&
    e^{-\im \frac{\phi_\mathbf{k}}{2}}
    )\mqty(a_\mathbf{k}\\ b_\mathbf{k}). 
\end{align}
Since the post-quench Hamiltonian~\eqref{eq:H} commutes with $\mathcal{P}$, the degree of inversion-symmetry breaking in the global state $\ket{\Psi_t}$ remains unchanged from that in the initial state $\ket{\Psi_0}$. 
Instead, by decomposing the whole system into a local subsystem $A$ and its complement $\Bar{A}$, we investigate the breaking of the inversion symmetry within the subsystem. 
As illustrated in Fig.~\ref{fig:subsystem}, we take as subsystem $A$ the stripe of length $\ell$ with the periodic boundary condition in the transverse direction. 
The subsystem is described by the reduced density matrix 
\begin{align}
\rho_A(t) = \Tr_{\bar{A}}(\ket{\Psi_t}\bra{\Psi_t}), 
\end{align}
where $\Tr_{\bar{A}}$ stands for the partial trace over subsystem $\bar{A}$. 
\par 
To study the inversion-symmetry breaking within subsystem $A$, we place the spatial origin at its center (denoted as a cross symbol in Fig.\ref{fig:subsystem}) so that the sites inside $A$ are mapped onto sites within $A$ under the inversion $\mathcal{P}$. 
We quantify the degree of the inversion-symmetry breaking within subsystem $A$ by using the R\'enyi entanglement asymmetry that is defined as~\cite{ares2023entanglement} 
\begin{align}
    \Delta S_A^{(n)}(t)
    = 
    \frac{1}{1-n}\ln(\frac{\Tr[\bar{\rho}_A(t)^n]}{\Tr[\rho_A(t)^n]}),
    \label{eq:REA}
\end{align}
where $n$ is the replica index and $\bar{\rho}_A=(\rho_A+\mathcal{P}\rho_A \mathcal{P}^{-1})/2$ is the symmetrization of $\rho_A$ with respect to the inversion $\mathcal{P}$. 
In the limit $n\to1$, Eq.~\eqref{eq:REA} reduces to the von-Neumann entanglement asymmetry, which is the Kullback-Leibler divergence between $\rho_A$ and $\bar{\rho}_A$, i.e., 
\begin{align}
    \Delta S_A^{(1)} (t)
    :=&\lim_{n\to1}\Delta S_A^{(n)}(t)
    \nonumber\\
    =& 
    \Tr[\rho_A(t)\{\log \rho_A(t)-\log \bar{\rho}_A(t)\}]. 
\end{align}
The  entanglement asymmetry has the desired properties as a measure of symmetry breaking~\cite{ma2022symmetric,han2023realistic}: it is positive semidefinite, $\Delta S_A^{(n)}\geq 0$, and reduces to zero if and only if the reduced density matrix is symmetric, i.e., $\Delta S_A^{(n)}=0\Leftrightarrow \mathcal{P}\rho_A\mathcal{P}^{-1}=\rho_A$. 

\section{Charged moments and dimensional reduction}\label{sec:dimensional reduction}
Here, we describe the method to calculate the  entanglement asymmetry by applying a dimensional reduction technique, which was originally introduced in Ref.~\cite{chung2000density} and has been employed to study the entanglement entropy of higher-dimensional systems both in and out of equilibrium~\cite{ares2014excited,murciano2020symmetry,yamashika2024time,frerot2016entanglement,frerot2015area}.
This approach exploits the translational invariance of the subsystem in one spatial direction to decompose the original two-dimensional problem into a set of independent one-dimensional ones, thereby making exact results for one-dimensional systems applicable. 
More recently, this framework has been extended to the entanglement asymmetry in two-dimensional square-lattice systems~\cite{yamashika2024entanglement,yamashika2025quenching}. 
Here, we apply it to the brickwork-lattice system corresponding to the honeycomb lattice.
\par
Substituting $\bar{\rho}_A=(\rho_A+\mathcal{P}\rho_A\mathcal{P}^{-1})/2$ into Eq.~\eqref{eq:REA}, the R\'enyi entanglement asymmetry can be written as 
\begin{align}
    \Delta S_A^{(n)}(t)
    =
    \frac{1}{1-n}\ln(\frac{1}{2^{n}} \sum_{\boldsymbol{\alpha}\in \qty{0,1}^n} 
    \frac{Z_n(\boldsymbol{\alpha},t)}{Z_n(\boldsymbol{0},t)}),
    \label{eq:REA_Z}
\end{align}
where $Z_n(\boldsymbol{\alpha},t)$ are the charged moments defined as 
\begin{align}
    Z_n(\boldsymbol{\alpha},t)
    =\Tr[\prod_{j=1}^n\rho_{A,\alpha_j}(t)], 
    \label{eq:charged moments}
\end{align}
with $\rho_{A,0}=\rho_A$ and $\rho_{A,1}=\mathcal{P}\rho_A \mathcal{P}^{-1}$. 
Note that, for $n=2$, the charged moments in Eq.~\eqref{eq:charged moments} reduces to the overlap between two density matrices, $Z_2(\boldsymbol{\alpha})=\Tr_{}(\rho_{A,\alpha_1}\rho_{A,\alpha_2})$, which is accessible in cold-atom experiments by using beam splitters and on-site occupation measurement~\cite{pichler2013thermal,cornfeld2019measuring,islam2015measuring,kaufman2016quantum}. 
\par 
Since the initial state $\ket{\Psi_0}$ is Gaussian and the post-quench Hamiltonian is quadratic, the time-evolved state $\ket{\Psi_t}$ and its reduced density matrix $\rho_A$ are also Gaussian and satisfy Wick's theorem. 
In addition, as the inversion $\mathcal{P}$ preserves the Gaussianity of the state, the inverted reduced density matrix $\rho_{A,1}=\mathcal{P}\rho_A\mathcal{P}^{-1}$ also satisfies Wick's theorem. 
This implies that $\rho_{A,\alpha}$ in Eq.~\eqref{eq:charged moments} can be univocally described by the two-point correlation matrix $\Gamma_\alpha$ whose entries are defined as~\cite{peschel2003calculation} 
\begin{align}
    [\Gamma_\alpha (t)]_{\bf i,i'}
    =
    2\Tr_A[\rho_{A,\alpha}(t) \mathbf{\Psi}_{\bf i} \mathbf{\Psi}_{\bf i'}^\dag]-\delta_{\mathbf{i,i'}}I,
    \label{eq:Gamma}
\end{align}
where $\mathbf{i,i'}\in A\cap \Lambda_{\mathrm{A}}$ and $\mathbf{\Psi_i}=(a_\mathbf{i},b_{\mathbf{i}+\mathbf{d}_{2}})^T$. Evaluating the correlators in Eq.~\eqref{eq:Gamma} with the time-evolved state $\mathcal{P}^\alpha\ket{\Psi_t}$, we obtain 
\begin{align}
    [\Gamma_\alpha(t)]_{\bf i,i'}
    = 
    \frac{1}{L_xL_y}
    \sum_\mathbf{k}
    e^{\im \mathbf{k\cdot(i-i')}}
    \mathcal{G}_{\mathbf{k},\alpha}(t), 
    \label{eq:Gamma_toeplitz}
\end{align}
where $\mathcal{G}_{\mathbf{k},\alpha}$ is the $2\times 2$ matrix defined as 
\begin{multline}
    \mathcal{G}_{\mathbf{k},\alpha}(t)
    = 
    e^{-\frac{\im}{2}(\mathbf{k\cdot d}_2+\phi_\mathbf{k})\sigma_z}
    [\sigma_x\sin \theta_\mathbf{k}\\
    +(-1)^\alpha\sigma_z \cos\theta_\mathbf{k}e^{2\im t \varepsilon_\mathbf{k}\sigma_x}]
    e^{\frac{\im}{2}(\mathbf{k\cdot d}_2+\phi_\mathbf{k})\sigma_z},
    \label{eq:Gamma_symbol}
\end{multline}
with $\sigma_{x,y,z}$ being the standard Pauli matrices. The reduced density matrix $\rho_{A,\alpha}$ can be expressed by using the two-point correlation matrix $\Gamma_{A,\alpha}$ as~\cite{peschel2003calculation}
\begin{multline}
    \rho_{A,\alpha}(t)
    =
    \det(\frac{I+\Gamma_\alpha(t)}{2})
    \\
    \times
    \exp(\sum_{\mathbf{i,i'}\in A\cap \Lambda_{\mathrm{A}}}\mathbf{\Psi}_{\bf i}^\dag \log(\frac{I-\Gamma_\alpha(t)}{I+\Gamma_\alpha(t)})_{\!\!\bf i,i'} \mathbf{\Psi}_{\bf i'}).
    \label{eq:rho_A}
\end{multline}
Plugging Eq.~\eqref{eq:rho_A} into Eq.~\eqref{eq:charged moments} and applying the Baker-Campbell-Hausdorff formula, we obtain 
\begin{multline}
    Z_n(\boldsymbol{\alpha},t)
    = 
    \det(\prod_{j=1}^n \frac{I+\Gamma_{\alpha_j}(t)}{2})
    \\
    \times\det(I+\prod_{j=1}^n \frac{I-\Gamma_{\alpha_j}(t)}{I+\Gamma_{\alpha_j}(t)})
    .
    \label{eq:Z_n}
\end{multline}
We give the derivation of Eq.~\eqref{eq:Z_n} in more detail in Appendix~\ref{app:derivation_chargedmoments}.  
\par 
Since the global state $\ket{\Psi_t}$ is invariant under translations and subsystem $A$ is taken to be periodic in the transverse direction, the reduced density matrix $\rho_A$ inherits the translational symmetry in that direction. 
To take advantage of this in the calculation of the charged moments in Eq.~\eqref{eq:Z_n}, we introduce the unitary matrix $U$ with entries 
\begin{align}
    U_{(i_x,i_y),(i_x',n_y)}
    = 
    \frac{\delta_{i_x,i_x'}e^{\frac{2\im \pi n_y}{L_y}i_y}}{\sqrt{L_y}}. 
    \label{eq:matrixU}
\end{align}
The matrix $U$ represents the partial Fourier transform only in the transverse direction. 
From Eqs.~\eqref{eq:Gamma_toeplitz} and \eqref{eq:matrixU}, we obtain 
\begin{align}
    U\Gamma_\alpha(t) U^\dag = \bigoplus_{n_y=0}^{L_y-1}
    \Gamma_{\alpha,k_y}(t). 
    \label{eq:Gamma_block}
\end{align}
Here, $\Gamma_{\alpha,k_y}$ is the $2\ell\times2\ell$ two-point correlation matrix for the one-dimensional system labeled by the transverse momentum $k_y=2\pi n_y/L_y$. In the thermodynamic limit in the longitudinal direction, $L_x\to\infty$, $\Gamma_{\alpha,k_y}$ reduces to the block-Toeplitz matrix as 
\begin{align}
    [\Gamma_{\alpha,k_y}(t)]_{i_x,i_x'}
    = 
    \int_{-\pi}^{\pi}
    \frac{dk_x}{2\pi}
    e^{\im k_x(i_x-i_x')}
    \mathcal{G}_{\mathbf{k},\alpha}(t), 
\end{align}
where the symbol $\mathcal{G}_{\mathbf{k},\alpha}$ is given in Eq.~\eqref{eq:Gamma_symbol}. 
Substituting identity $I=UU^\dag=U^\dag U$ in Eq.~\eqref{eq:Z_n} and employing the block-diagonal structure~\eqref{eq:Gamma_block}, we can decompose the charged moments into the product of independent contributions of each transverse-momentum sector as 
\begin{align}
    Z_n(\boldsymbol{\alpha},t)
    =\prod_{n_y=0}^{L_y-1} Z_{n,k_y}(\boldsymbol{\alpha},t),
    \label{eq:Z_product}
\end{align}
where 
\begin{multline}
    Z_{n,k_y}(\boldsymbol{\alpha},t)
    = 
    \det(\prod_{j=1}^n \frac{I+\Gamma_{\alpha_j,k_y}(t)}{2})
    \\
    \times\det(I+\prod_{j=1}^n \frac{I-\Gamma_{\alpha_j,k_y}(t)}{I+\Gamma_{\alpha_j,k_y}(t)}).
    \label{eq:Z_nk}
\end{multline}
Note that $Z_{n,k_y}(\boldsymbol{\alpha},t)$ in the above equation are the charged moments for a one-dimensional state labeled by the transverse momentum $k_y=2\pi n_y/L_y$.  
In the following sections, we derive analytical expressions for the entanglement asymmetry in the ground state and the time-evolved state by combining well-developed techniques for one-dimensional systems with Eq.~\eqref{eq:Z_nk}. 

\section{Entanglement asymmetry in the ground state of pre-quench Hamiltonian}
\label{sec:result_GS}
In this section, we investigate the entanglement asymmetry in the ground state of the Hamiltonian \eqref{eq:H_0}, which serves as the initial state of our quench protocol.
In particular, we analytically calculate the asymptotic form of the entanglement asymmetry for $\ell\gg1$ taking the thermodynamic limit in the longitudinal direction, $L_x\to\infty$. 
\par 
As shown in Sec.~\ref{sec:dimensional reduction}, the R\'enyi entanglement asymmetry for the ground state can be calculated from the momentum-resolved charged moments in Eq.~\eqref{eq:Z_nk} at $t=0$. 
For clarity, we first present the calculation of the charged moments for $n=2$. In this case, Eq.~\eqref{eq:Z_nk} at $t=0$ simplifies as
\begin{align}
    Z_{2,k_y}(\boldsymbol{\alpha},0)
    = 
    \det(\frac{I+\Gamma_{\alpha_1,k_y}(0)\Gamma_{\alpha_2,k_y}(0)}{2}).
    \label{eq:Z_2k(t=0)}
\end{align}
The above expression involves the product of the block-Toeplitz matrices, $\Gamma_{\alpha_1,k_y}(0)\Gamma_{\alpha_2,k_y}(0)$, which is in general not a block-Toeplitz matrix. However, for $\ell\gg1$, the product of block-Toeplitz matrices can be approximated by the block-Toeplitz matrix generated by the product of the symbols of the factors, see e.g. Ref.~\cite{ares2023lack}. This allows us to approximate Eq.~\eqref{eq:Z_2k(t=0)} as 
\begin{align}
    Z_{2,k_y}(\boldsymbol{\alpha},0)
    \simeq 
    \det \mathrm{T}[z_{2,\mathbf{k}}(\boldsymbol{\alpha})]. 
    \label{eq:Z_2k_apprx}
\end{align}
Here, $\mathrm{T}[z_{2,\mathbf{k}}(\boldsymbol{\alpha})]$ is the $2\ell\times 2\ell$ block-Toeplitz matrix whose elements are given by 
\begin{align}
    \mathrm{T}[z_{2,\mathbf{k}}(\boldsymbol{\alpha})]_{i_xi_x'}
    =
    \int_{-\pi}^\pi
    \frac{dk_x}{2\pi}
    e^{\im  k_x(i_x-i_x')}
    z_{2,\mathbf{k}}(\boldsymbol{\alpha}),
\end{align}
with the $2\times 2$ symbol
\begin{align}
    z_{2,\mathbf{k}}(\boldsymbol{\alpha})
    =
    \frac{I+\mathcal{G}_{\mathbf{k},\alpha_1}(0)\mathcal{G}_{\mathbf{k},\alpha_2}(0)}{2}. 
\end{align}
The asymptotic form of the determinant in Eq.~\eqref{eq:Z_2k_apprx} for $\ell\gg1$ can be evaluated using Widom-Szeg\H{o} theorem~\cite{widom1974asymptotic}, which results in
\begin{align}
    \ln\frac{Z_{2,k_y}(\boldsymbol{\alpha},0)}{Z_{2,k_y}(\boldsymbol{0},0)}
    &\simeq 
    \ell \int_{-\pi}^{\pi}\frac{dk_x}{2\pi}
    \ln \frac{\det z_{2,\mathbf{k}}(\boldsymbol{\alpha})}{\det z_{2,\mathbf{k}}(\boldsymbol{0})}
    \\
    &=
    \ell|\alpha_1-\alpha_2| \int_{-\pi}^{\pi}\frac{dk_x}{2\pi}\ln \sin^2\theta_\mathbf{k}.
    \label{eq:Z_2k_asymptotic}
\end{align}
As shown in Appendix~\ref{app:n-thEA_GS}, Eq.~\eqref{eq:Z_2k_asymptotic} can be generalized to $n\geq 2$ as 
\begin{align}
\ln\frac{Z_{n,k_y}(\boldsymbol{\alpha},0)}{Z_{n,k_y}(\boldsymbol{0},0)}
\simeq 
\ell N^{(n)}(\boldsymbol{\alpha})  \int_{-\pi}^\pi \frac{dk_x}{4\pi}\ln \sin^2\theta_\mathbf{k}, 
\label{eq:Z_nk/Z_nk}
\end{align}
with $N^{(n)}(\boldsymbol{\alpha})=\sum_{j=1}^n|\alpha_j-\alpha_{j+1}|$ and $\alpha_{n+1}=\alpha_1$. 
Applying it to Eq.~\eqref{eq:Z_product}, we obtain the charged moments 
\begin{align}
    \frac{Z_{n}(\boldsymbol{\alpha},0)}{Z_{n}(\boldsymbol{0},0)}
    \simeq 
    X_0^{N_n(\boldsymbol{\alpha})},
    \label{eq:Z/Z_n}
\end{align}
where 
\begin{align}
     X_0
     =
     \exp(
     -\ell 
     \sum_{n_y=0}^{L_y-1}
     \int_{-\pi}^{\pi}
     \frac{dk_x}{4\pi}
     \ln(1+\frac{M^2}{\varepsilon_\mathbf{k}^2})). 
     \label{eq:J_0}
\end{align} 
Here, we used $\sin^2\theta_\mathbf{k}=\varepsilon_\mathbf{k}^2/(\varepsilon_\mathbf{k}^2+M^2)$ from Eq.~\eqref{eq:angles}. 
\par 
Substituting Eq.~\eqref{eq:Z/Z_n} into Eq.~\eqref{eq:REA_Z}, we obtain the R\'enyi entanglement asymmetry. 
The summation over $\boldsymbol{\alpha}\in\qty{0,1}^n$ in Eq.~\eqref{eq:REA_Z} can be calculated analytically, as described in Appendix~\ref{app:summation alpha}. 
We finally arrive at the concise expression for the R\'enyi entanglement asymmetry of the ground state, 
\begin{align}
    \Delta S_A^{(n)}(0)
    \simeq 
    H^{(n)}(X_0),
    \label{eq:REA_t=0}
\end{align}
where 
\begin{align}
    H^{(n)}(x)=\frac{1}{1-n}\ln(\qty[\frac{1+x}{2}]^n+\qty[\frac{1-x}{2}]^n)
    \label{eq:H^n}
\end{align}
denotes the R\'enyi entropy of order $n$ for a single bit of information.
Taking the limit $n\to1$ in Eq.~\eqref{eq:REA_t=0}, we also obtain the von-Neumann entanglement asymmetry for the ground state as 
\begin{align}
    \Delta S_1^{(1)}(0)
    \simeq 
    H^{(1)}(X_0),
    \label{eq:EA_t=0}
\end{align}
with 
\begin{align}
    H^{(1)}(x):=&\lim_{n\to1}H^{(n)}(x) \nonumber
    \\
    =&
    -\frac{1+x}{2}\ln\frac{1+x}{2}
    -\frac{1-x}{2}\ln\frac{1-x}{2}.
\end{align}

\begin{figure}
    \raggedright
    \includegraphics[width=0.95\linewidth]{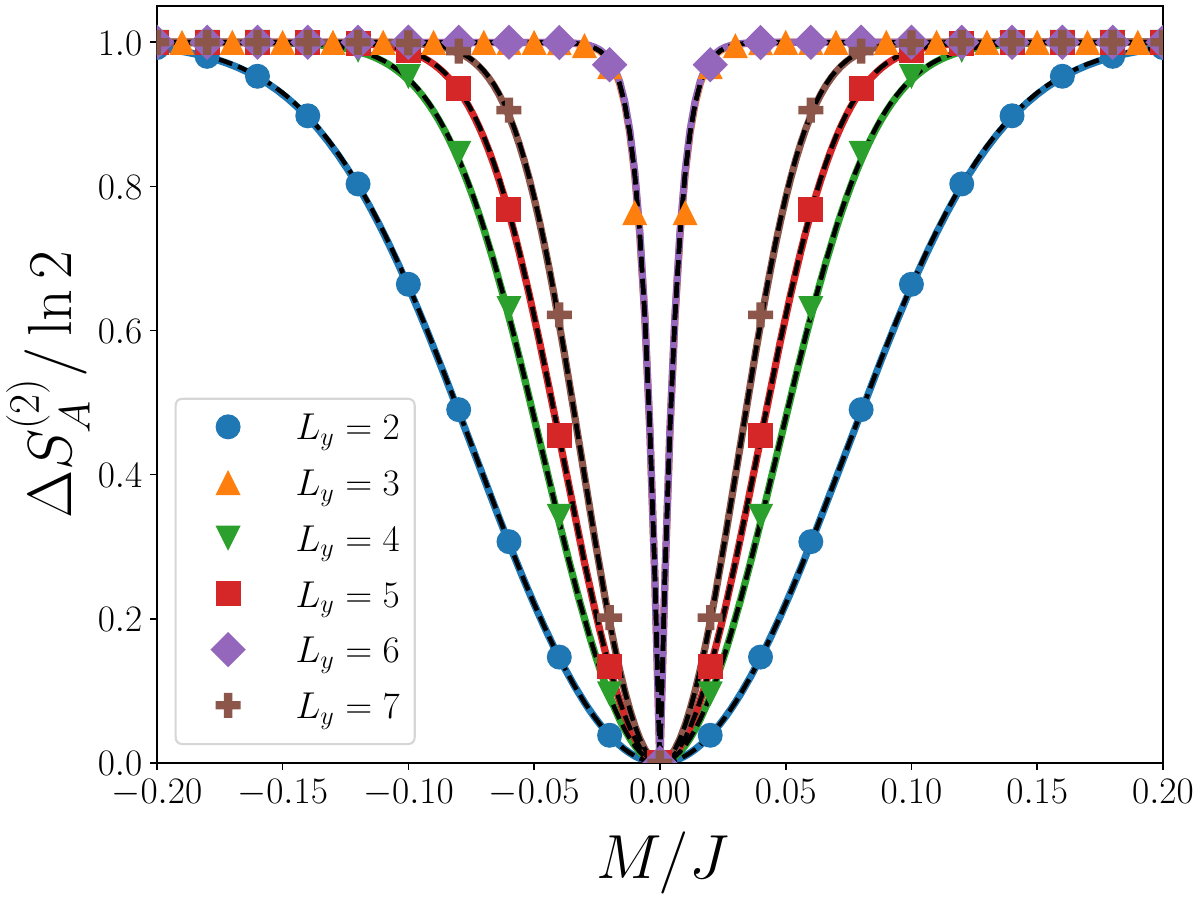}
    \caption{The $n=2$ R\'enyi entanglement asymmetry of the ground state~\eqref{eq:Psi_0}. The solid curves and symbols are the analytical result in Eq.~\eqref{eq:REA_t=0} with $X_0$ obtained using Eq.~\eqref{eq:J_0} and the exact results obtained numerically by evaluating Eq.~\eqref{eq:Z_nk}, respectively. The black dashed curves denote the asymptotic approximations of Eq.~\eqref{eq:REA_t=0} with Eqs.~\eqref{eq:J_0_1} and \eqref{eq:J_0_2}. We take $\ell=100$ for all the plots.} 
    \label{fig:REA_t=0}
\end{figure}
\par 
In Fig.~\ref{fig:REA_t=0}, we show the $n=2$ R\'enyi entanglement asymmetry of the ground state as a function of $M$.
It shows that the analytical result in Eq.~\eqref{eq:REA_t=0} agrees excellently with the exact ones obtained by numerically evaluating the momentum-resolved charged moments in Eq.~\eqref{eq:Z_nk}. 
As expected, the entanglement asymmetry decreases as $|M/J|$ gets smaller and vanishes at the symmetric point $M/J=0$. 
For large $|M/J|$, the  entanglement asymmetry saturates to the maximal value $\ln 2$. This saturation is a general property of entanglement asymmetry for discrete symmetries reported in Ref.~\cite{ferro2024non}, which shows that for a $\mathbb{Z}_N$ symmetry the entanglement asymmetry saturates to $\ln N$ for $\ell\gg1$. 
Indeed, since $X_0$ in Eq.~\eqref{eq:J_0} decays exponentially with respect to $\ell$, we obtain from Eq.~\eqref{eq:REA_t=0} that 
\begin{align}
\lim_{\ell\to\infty} \Delta S_A^{(n)}(0)= H^{(n)}(0)=\ln2, 
\label{eq:S=ln2}
\end{align}
for $M/J\neq0$. 
\par
We also find in Fig.~\ref{fig:REA_t=0} that the entanglement asymmetry exhibits nonanalytic behavior at $M=0$ and becomes independent of $L_y$ when $L_y$ is divisible by three.
Otherwise, it is smooth around $M/J=0$ and increases with $L_y$. 
This peculiar dependence in the behavior of the entanglement asymmetry on $L_y$ is attributed to the presence of Dirac points in the Brillouin zone: 
According to Eq.~\eqref{eq:REA_t=0}, the entanglement asymmetry in the ground state is univocally described by $X_0$, which is given in Eq.~\eqref{eq:J_0}. 
When $L_y$ is not divisible by three, the quantized momenta $\mathbf{k}$ cannot take the Dirac points and hence $\varepsilon_\mathbf{k}>0~\forall \mathbf{k}$. 
In this case, the integrand on the right-hand side of Eq.~\eqref{eq:J_0} for small $M/J$ can be expanded in terms of $M/\varepsilon_\mathbf{k}$ as
\begin{align}
X_0 &\simeq \exp(-\frac{\ell M^2}{4\pi}\sum_{n_y=0}^{L_y-1}\int_{-\pi}^\pi\frac{dk_x}{\varepsilon_\mathbf{k}^2}).\label{eq:J_0_1}
\end{align}
The above result shows that the entanglement asymmetry in Eq.~\eqref{eq:REA_t=0} depends on $L_y$ and is analytic around $M/J=0$. 
On the other hand, when $L_y$ is divisible by three, $\varepsilon_\mathbf{k}$ vanishes at the Dirac points $\mathbf{k}=\pm(2\pi/3,4\pi/3)$. 
In this case, since the integrand in Eq.~\eqref{eq:J_0} diverges at $\varepsilon_\mathbf{k}=0$, the summation over $n_y$ is dominated by the terms with $n_y=L_y/3$ and $2L_y/3$, on which the Dirac points appear. Equation~\eqref{eq:J_0} can then be approximated as
\begin{align}
X_0
\simeq
\exp(-\ell \sum_{k_y=\pm 2\pi/3}\int_{-\pi}^\pi \frac{dk_x}{4\pi}\ln(1+\frac{M^2}{\varepsilon_\mathbf{k}^2})).  
\label{eq:J_0_2_1}
\end{align}
Observing that the main contribution of the integral in the above equation comes from the vicinity of the Dirac points where $\varepsilon_\mathbf{k}\simeq J|k_x\mp 2\pi/3|$ with fixed $k_y=\pm 2\pi/3$, Eq.~\eqref{eq:J_0_2_1} can be evaluated as 
\begin{align}
    X_0\simeq 
    \exp(-\ell \int_\mathbb{R} \frac{dk_x}{2\pi} \ln (1+\frac{M^2}{J^2k_x^2}))
    =
    e^{-\frac{\ell |M|}{J}}.
    \label{eq:J_0_2}
\end{align}
Here, we extended the integral domain to $k_x\in \mathbb{R}$ using the fact that the integrand in the above expression is suppressed away from the Dirac points. 
Equation~\eqref{eq:J_0_2} clearly demonstrates that the entanglement asymmetry in Eq.~\eqref{eq:REA_t=0} exhibits nonanalytic behavior at $M=0$ and is independent of $L_y$. 
The entanglement asymmetry evaluated with Eqs.~\eqref{eq:J_0_1} and \eqref{eq:J_0_2} is plotted by the black dashed lines in Fig.~\ref{fig:REA_t=0}, showing an excellent agreement with the exact numerical result.

\section{Quench dynamics of entanglement asymmetry}
\label{sec:result_TEV}
\begin{figure*}
    \centering
    \includegraphics[width=\linewidth]{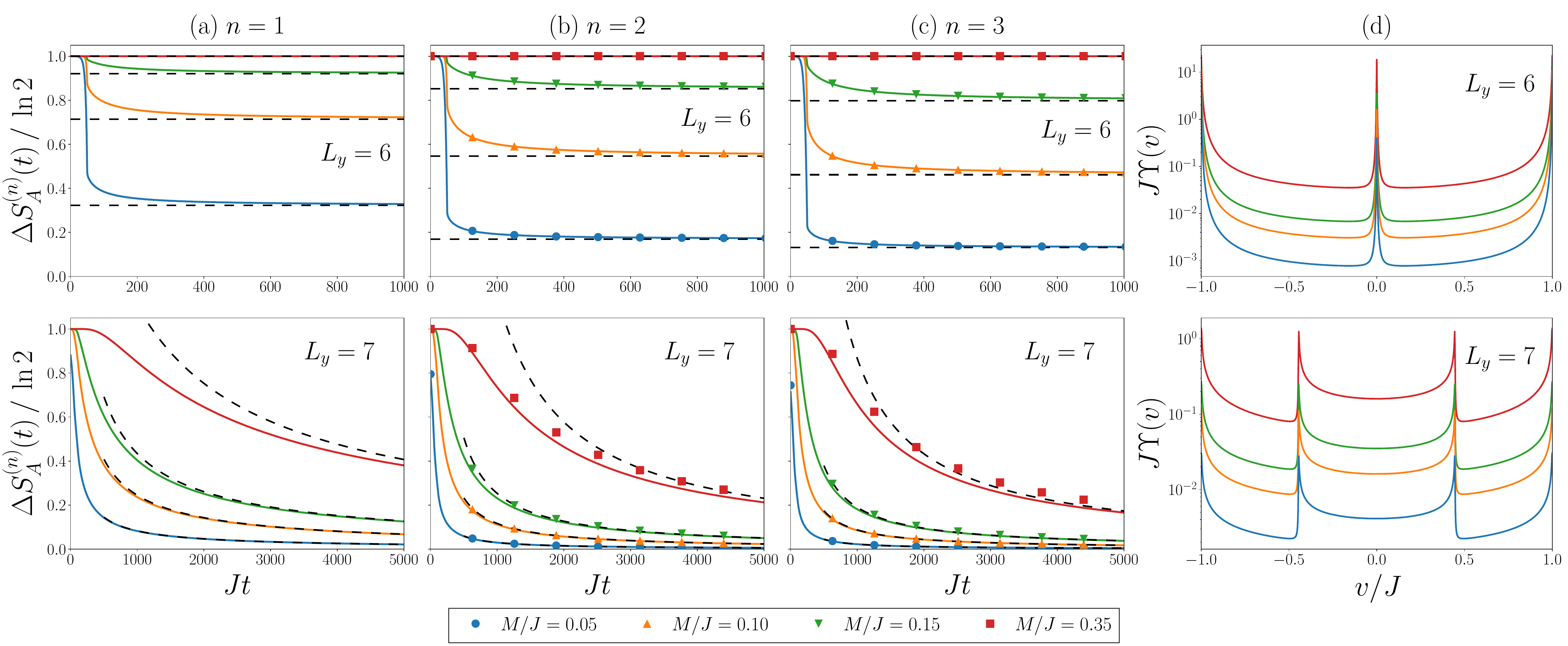}
    \caption{
    (a)-(c) 
    Time evolution of the $n$-th order R\'enyi entanglement asymmetry after the quench into the Hamiltonian~\eqref{eq:H} starting from the ground state of the Hamiltonian~\eqref{eq:H_0} for several finite $M/J$ values. 
    The solid curves denote the analytical result in Eq.~\eqref{eq:REA_t} with Eq.~\eqref{eq:J_z}. 
    The symbols are the exact value of the R\'enyi entanglement asymmetry obtained by numerically evaluating the charged moments using Eq.~\eqref{eq:Z_nk}. 
    The dashed lines correspond to Eqs.~\eqref{eq:REA_large t} and \eqref{eq:REA_inf} that predict the asymptotic form of the entanglement asymmetry at large times. 
    We take $\ell=100$ in all the plots. 
    (d) 
    $\Upsilon (v)$ defined in Eq.~\eqref{eq:Ups} that characterizes how much the quasiparticle pairs with the longitudinal group velocity $v$ contribute to the entanglement asymmetry. 
    To numerically evaluate $\Upsilon(v)$, we approximate the Dirac delta function in Eq.~\eqref{eq:Ups} by the Lorentz function $\delta_\eta(x)=\eta/[\pi(x^2+\eta^2)]$ with $\eta=10^{-3}$.}  
    \label{fig:REA_TEV}
\end{figure*}
In this section, we investigate the time evolution of the inversion-symmetry breaking after the quantum quench starting from $M\neq 0$ into $M=0$. 
We analytically derive the asymptotic form of the  entanglement asymmetry in the ballistic limit, $Jt,\ell\to\infty$ with $Jt/\ell$ fixed, by taking the thermodynamic limit in the longitudinal direction $L_x\to\infty$. 
\par 
Using the fact that Eqs.~\eqref{eq:REA_t=0} and \eqref{eq:EA_t=0} are derived by decomposing the charged moments into independent one-dimensional contributions labeled by $k_y = 2\pi n_y/L_y$, we can extend it to finite times by invoking the quasiparticle picture, which is a semiclassical description for the quench dynamics of one-dimensional integrable systems~\cite{calabrese2005evolution,fagotti2008evolution,alba2017entanglement}.
According to this picture, a global quench at $t=0$ uniformly creates quasiparticle excitations throughout the system, and those generated at the same position form entangled pairs that ballistically propagate in time. 
The ratio $Z_{n,k_y}(\boldsymbol{\alpha},t)/Z_{n,k_y}(\boldsymbol{0},t)$ is then determined by the number of the pairs that remain inside the subsystem, while the integrand $\ln \sin^2 \theta_\mathbf{k}$ in Eq.~\eqref{eq:Z_nk/Z_nk} represents the contribution of each quasiparticle mode to the charged moments~\cite{rylands2024microscopic,ares2023entanglement,murciano2024entanglement,caceffo2024entangled}. 
The integrand can be expressed by using the correlation function between quasiparticle excitations as 
\begin{align}
    \ln \sin^2 \theta_\mathbf{k}
    = 
    \ln_{}(1-4|\bra{\Psi_0}\gamma_\mathbf{k+}^\dag \gamma_\mathbf{k-}\ket{\Psi_0}|^2). 
    \label{eq:ln(sin)}
\end{align}
Equation~\eqref{eq:ln(sin)} suggests that, in our setting, the quasiparticles in the upper and lower bands are excited as the pairs by the quench. 
Since their dispersion relations are $\pm \varepsilon_\mathbf{k}$, their group velocities are opposite, $\pm \partial_{k_x}\varepsilon_\mathbf{k}$.
As a result, the pairs propagate in the opposite direction after the quench, and the number of the pairs inside the subsystem gradually decreases in time. 
This effect can be incorporated by replacing the prefactor $\ell$ in Eq.~\eqref{eq:Z_nk/Z_nk} with $\max(\ell-2|\partial_{k_x}\varepsilon_\mathbf{k}|t,0)$~\cite{rylands2024microscopic,ares2023entanglement,murciano2024entanglement,caceffo2024entangled}. 
We then obtain the asymptotic form of the momentum-resolved charged moments in the ballistic limit, $Jt,\ell\to\infty$ with $Jt/\ell$ fixed, as 
\begin{align}
    \ln
    \frac{Z_{n,k_y}(\boldsymbol{\alpha},t)}{Z_{n,k_y}(\boldsymbol{0},t)}
    \simeq 
    \ell N^{(n)}(\boldsymbol{\alpha}) \int_{-\pi}^\pi \frac{dk_x}{4\pi} x_\zeta(\mathbf{k} )\ln \sin^2\theta_\mathbf{k}, 
    \label{eq:Z/Zn(t)}
\end{align}
where $x_\zeta(\mathbf{k})=\max(1-2|\partial_{k_x} \varepsilon_\mathbf{k}|\zeta,0)$ denotes the fraction of the pairs with momentum $\mathbf{k}$ that are inside the subsystem at the rescaled time $\zeta=t/\ell$. 
\par 
Using Eq.~\eqref{eq:Z/Zn(t)} and following the same calculation as those in deriving Eq.~\eqref{eq:REA_t=0}, we obtain the asymptotic form of the entanglement asymmetry after the quench, 
\begin{align}
    \Delta S_A^{(n)}(t)
    \simeq 
    H^{(n)}(X_\zeta),
    \label{eq:REA_t}
\end{align}
where
\begin{align}
    X_\zeta
    =
    \exp(\ell \sum_{n_y=0}^{L_y-1} \int_{-\pi}^{\pi}\frac{dk_x}{4\pi} x_\zeta(\mathbf{k})\ln \sin^2 \theta_\mathbf{k}).
    \label{eq:J_z}
\end{align}
In Figs.~\ref{fig:REA_TEV} (a)-(c), we show the time evolution of the entanglement asymmetry after the quench from the ground state at $M\neq0$ to $M=0$ with several fixed values of $L_y$ and the replica index $n$. Our analytical result in Eq.~\eqref{eq:REA_t} agrees excellently with  the exact numerical results obtained by using Eq.~\eqref{eq:Z_nk}.
Notably, the qualitative behavior of the R\'enyi entanglement asymmetry does not depend on the replica index $n$. 
\par 
The lower panels of Figs.~\ref{fig:REA_TEV} (a)-(c) show that, for odd $L_y$, the entanglement asymmetry tends to zero, suggesting that the inversion symmetry broken in the initial state is restored after the quench. 
This is consistent with the expectation that the reduced density matrix $\rho_A$ for the subsystem relaxes into its corresponding GGE, $\rho_A(t\to\infty)=\Tr_{\bar{A}}(\rho_{\rm GGE})$~\cite{rigol2007relaxation,caux2013time,ilievski2015complete,vidmar2016generalized,essler2016quench,langen2015experimental,barthel2008dephasing,cramer2008exact,calabrese2012quantum}, where
\begin{align}
    \rho_{\rm GGE}
    =\frac{e^{\sum_{\mathbf{k},s=\pm}\lambda_{\mathbf{k},s}\gamma_{\mathbf{k}s}^\dag \gamma_{\mathbf{k}s}}}{\Tr[e^{\sum_{\mathbf{k},s=\pm}\lambda_{\mathbf{k},s}\gamma_{\mathbf{k}s}^\dag \gamma_{\mathbf{k}s}}]}.
    \label{eq:GGE}
\end{align}
Here, $\{\lambda_{\mathbf{k},\pm}\}$ are the Lagrange multipliers determined by the conservation laws,
\begin{align}
    \Tr[\rho_{\rm GGE}\gamma_{\mathbf{k}\pm}^\dag \gamma_{\mathbf{k}\pm}]
    =
    \bra{\Psi_0}\gamma_\mathbf{k\pm}^\dag \gamma_\mathbf{k\pm}\ket{\Psi_0}.
\end{align}
One readily finds that the GGE respects the inversion symmetry, i.e., $\mathcal{P}\rho_{\rm GGE}\mathcal{P}^{-1}=\rho_{\rm GGE}$. 
Therefore, as the reduced density matrix approaches the GGE, the inversion symmetry broken in the initial state is restored, and the entanglement asymmetry vanishes in the large-time limit. On the other hand, we observe in the upper panels of Figs.~\ref{fig:REA_TEV} (a)-(c) that, when $L_y$ is even, the entanglement asymmetry remains finite even at large times, indicating the absence of symmetry restoration. 
This result suggests that the reduced density matrix does not relax into the GGE in Eq.~\eqref{eq:GGE}. 
We have thus found, both analytically and numerically, that the manner in which the subsystem relaxes critically depends on its geometry.

\section{Absence of inversion-symmetry restoration} 
\label{sec:lack}
In the previous section, we found that the behavior of the entanglement asymmetry after the quench drastically changes depending on the parity of the subsystem size in the transverse direction. 
In particular, when $L_y$ is even, the entanglement asymmetry does not tend to zero even in the large-time limit $t\to\infty$. That is, the inversion symmetry broken by the initial state is not restored after the quench. 
In this section, we clarify the physical origin of this anomalous behavior based on the quasiparticle picture.
\par 
As shown in Eq.~\eqref{eq:REA_t}, the entanglement asymmetry after the quench is given by $X_\zeta$.
To deduce the behavior of the entanglement asymmetry at large times, we first rewrite $X_\zeta$ in Eq.~\eqref{eq:J_z} as
\begin{align}
    X_\zeta=\exp(-\ell \int_{\mathbb{R}}dv \tilde{x}_\zeta(v)\Upsilon(v)),
    \label{eq:J_z(v)}
\end{align}
where
\begin{align}
    \Upsilon(v)=-\sum_{n_y=0}^{L_y-1}\int_{-\pi}^{\pi}
    \frac{dk_x}{4\pi}\delta(v-\partial_{k_x}\varepsilon_\mathbf{k})\ln\sin^2\theta_\mathbf{k}. 
    \label{eq:Ups}
\end{align}
Here, $\tilde{x}_\zeta(v)=\max(1-2|v|\zeta,0)$ denotes the fraction of quasiparticle pairs with longitudinal velocities $ \pm \partial_{k_x} \varepsilon_\mathbf{k}=\pm v$ that remain inside the subsystem at the rescaled time $\zeta=t/\ell$, while $\Upsilon(v)$ quantifies the weight of their contribution to the entanglement asymmetry. 
At time $\zeta$, $\tilde{x}_\zeta(v)$ vanishes for $2|v|\zeta>1$, thereby filtering out the contributions of fast-moving quasiparticle pairs that have already escaped from the subsystem. 
The entanglement asymmetry at large times $\zeta\gg1$ is therefore governed solely by the slow-moving quasiparticle pairs, whose contributions are given by $\Upsilon(v)$ in the vicinity of $v=0$. 
\par 
In Fig.~\ref{fig:REA_TEV} (d), we show the function $\Upsilon(v)$ for the initial states considered in the other panels. 
To numerically evaluate $\Upsilon(v)$, we replaced the Dirac delta function in Eq.~\eqref{eq:Ups} with the Lorentz function $\delta_\eta(x)=\eta/[\pi(x^2 +\eta^2)]$, which reduces to $\delta(x)$ in the limit $\eta\to0$. 
These figures show that $\Upsilon(v)$ exhibits a sharp peak at $v=0$ when $L_y$ is even, whereas this peak is absent otherwise. 
The height of the peak at $v=0$ diverges as $\Upsilon(0)\propto \eta^{-1}$, implying that, for even $L_y$, a macroscopic number of quasiparticle pairs with zero longitudinal velocity are excited by the quench.  
Since these pairs never leave the subsystem, the entanglement asymmetry does not tend to zero and the inversion symmetry is not restored, as observed in the upper panels of Figs.~\ref{fig:REA_TEV}~(a)-(c).
\par 
The divergent peak of $\Upsilon(v)$ at $v=0$ originates from the fact that, as seen in Fig.~\ref{fig:band}~(a), the dispersion relation $\varepsilon_\mathbf{k}$ becomes independent of $k_x$ at $k_y=\pi$, which occurs when $L_y$ is even.
In other words, all $L_x$ quasiparticle modes with fixed $k_y=\pi$ have zero longitudinal group velocity, resulting in an infinite number of such modes in the thermodynamic limit $L_x\to\infty$.  
Consequently, their total contribution to the entanglement asymmetry, $\Upsilon(0)$, diverges. 
We thus find that the absence of inversion-symmetry restoration originates from the presence of the flat energy dispersion with fixed $k_y=\pi$, elucidating the critical role of the band structure and its exotic dispersion on the symmetry restoration. 
This mechanism of the absence of symmetry restoration is distinct from those reported in previous studies, such as the activation of non-Abelian charges~\cite{murciano2024entanglement} or the Bose-Einstein condensation~\cite{yamashika2025quenching}. 
\par 
Although we have focused on the sudden quench dynamics from the half-filled ground state for simplicity, the above mechanism does not rely on half filling. Away from half filling, one evaluates the correlation matrix with the initial ground state at the chosen chemical potential, which changes quasiparticle mode occupations quantitatively but does not alter the basic role of the zero-longitudinal-velocity modes. In particular, whenever the post-quench Hamiltonian exhibits a flat band, the sudden quench is expected to excite an extensive set of modes with zero longitudinal group velocity, so that the entanglement asymmetry remains finite in the long-time limit. In contrast, if the change of $M$ is not sudden, the excitation of these zero-velocity quasiparticles can be adiabatically suppressed once the ramp time exceeds the inverse of the gap between the flat bands ($\sim 1/(2J)$, see Fig.~\ref{fig:band} (a)), potentially mitigating the absence of symmetry restoration.
\par 
In closing this section, we derive explicit expressions for the behavior of the entanglement asymmetry at large times.
As argued above, this is governed by $\Upsilon(v)$ in the vicinity of $v=0$. 
For odd $L_y$, expanding $\Upsilon(v)$ in Eq.~\eqref{eq:J_z(v)} in terms of $v$ and performing the integral over $v$, we obtain 
\begin{align}
    X_\zeta\simeq1-\frac{\ell \Upsilon(0)}{2\zeta}+O(\zeta^{-2}). 
\end{align}
Substituting the above expression into Eq.~\eqref{eq:REA_t} and expanding it in terms of $\zeta^{-1}$, we obtain 
\begin{align}
    \Delta S_A^{(n)}(t)
    \simeq 
    \frac{\ell \Upsilon(0)}{4\zeta}
    \times 
    \begin{dcases}
        \frac{n}{n-1} &n\geq2,\\
        \ln\frac{4e\zeta}{\Upsilon(0)\ell} & n\to1
    \end{dcases},
    \label{eq:REA_large t}
\end{align}
for $Jt\gg\ell$. 
The above expression shows that the entanglement asymmetry tends to zero as $(Jt)^{-1}$ [$(Jt)^{-1}\ln (Jt)$ when $n\to1$], indicating that the inversion symmetry broken in the initial state is restored after the quench. 
In the lower panels of Figs.~\ref{fig:REA_TEV} (a)-(c), we see that Eq.~\eqref{eq:REA_large t} shown as dashed curves well reproduces the long-time asymptotic behavior of the entanglement asymmetry.
\par 
For even $L_y$, the function $x_{\zeta}(\mathbf{k})$ in Eq.~\eqref{eq:J_z} vanishes except for $k_y=\pi$, where it reduces to unity because $\partial_{k_x} \varepsilon_\mathbf{k}|_{k_y=\pi}=0$, at large times. 
This allows us to replace $x_{\zeta}(\mathbf{k})$ in Eq.~\eqref{eq:J_z} with $\delta_{k_y,\pi}$, which results in
\begin{align}
    X_{\zeta}
    \simeq[1+(M/J)^2]^{-\ell/2}+O(\zeta^{-1}).
    \label{eq:J_inf}
\end{align}
Here, we used $\varepsilon_\mathbf{k}|_{k_y=\pi}=J$. 
Substituting Eq.~\eqref{eq:J_inf} into Eq.~\eqref{eq:REA_t}, we obtain the saturation value of the entanglement asymmetry in the large-time limit as 
\begin{align}
\lim_{t\to\infty}\Delta S_A^{(n)}(t) = H^{(n)}([1+(M/J)^2]^{-\ell/2}).  
\label{eq:REA_inf}
\end{align}
Note that the above saturation value is independent of $L_y$, implying that the contribution of the zero-velocity modes remains finite for arbitrarily large even $L_y$.
As shown in the upper panels of Figs.~\ref{fig:REA_TEV} (a)-(c), Eq.~\eqref{eq:REA_inf} excellently agrees with the saturation value of the entanglement asymmetry.
Furthermore, combining Eq.~\eqref{eq:REA_inf} with Eq.~\eqref{eq:S=ln2}, we obtain
\begin{align}
    \lim_{\ell\to\infty}\Delta S_A^{(n)}(0)=\lim_{\ell\to\infty}\Delta S_A^{(n)}(\infty)=\ln2. 
\end{align}
This means that, despite the symmetric dynamics, the inversion symmetry broken in the initial state remains intact throughout the time evolution, owing to the macroscopic occupation of quasiparticle modes with zero longitudinal velocity.
\par 
As shown by the red squares in the upper panels of Figs.~\ref{fig:REA_TEV}~(a)-(c), this effect is visible even for finite $\ell$. This suggests that the absence of symmetry restoration can be observed even for a relatively small-sized system, opening the door for testing our prediction in cold atom experiments~\cite{tarruell2012creating,uehlinger2013artificial,jotzu2014experimental}.
\par

\section{Conclusion}\label{sec:conclusion}
We have investigated the space-inversion symmetry within a subsystem of free fermions on a honeycomb lattice.
By computing the entanglement asymmetry analytically and numerically, we demonstrated that the subsystem geometry and the band structure play crucial roles in symmetry breaking both in and out of equilibrium.
\par 
In the ground state, the entanglement asymmetry exhibits a nonanalytic dependence on the sublattice imbalance $M$ when the transverse subsystem size $L_y$ is multiple of three, where the quantized momenta can take the Dirac points. For the quench dynamics from $M\neq0$ to $M=0$, the parity of $L_y$ governs whether the inversion symmetry is restored or remains broken: For odd $L_y$, the inversion symmetry is restored as the subsystem relaxes into the symmetric GGE. In contrast, for even $L_y$, a flat dispersion with a fixed transverse momentum leads to a macroscopic occupation of quasiparticle modes with zero group velocity, which prevents the subsystem from relaxing and restoring the symmetry. 
\par 
We stress that the above phenomena are governed not only by the bulk band structure, but also by the system geometry and boundary conditions, which determine a discrete set of allowed momenta. Therefore, for other lattice geometries, qualitatively similar results are expected when the corresponding quantized momenta contain the Dirac points and support an extensive set of zero-velocity modes.
\par 
Our quench protocol can be realized in cold atoms in optical lattices~\cite{tarruell2012creating,uehlinger2013artificial,jotzu2014experimental}, where both the lattice geometry and the energy imbalance can be precisely controlled. 
In such setups, the $n=2$ R\'enyi entanglement asymmetry is expected to be experimentally accessible by combining the beam-splitter interference with the site-resolved imaging, which may enable the direct measurement of the charged moment~\cite{pichler2013thermal,cornfeld2019measuring,islam2015measuring,kaufman2016quantum}. 
Experimentally verifying our predictions would provide valuable insights into the connections between relaxation dynamics, system geometry, and band structure.

\acknowledgements 
We thank Filiberto Ares and Soma Takemori for fruitful discussions. 
SY was supported by Grant-in-Aid for Young Scientists (Start-up) No.\,25K23355 and Institute for Advanced Science, University of Electro-Communications. SE acknowledges support from JSPS KAKENHI Grant Numbers JP23H01174,
JP25K00217, Matsuo Foundation, and Institute for Advanced Science, University of Electro-Communications.

\section*{Data availability}
The data the support the findings of this article are openly available 
\cite{Hara2026_HoneycombEA}.

\appendix
\section{Derivation of Eq.~\eqref{eq:Z_n}}
\label{app:derivation_chargedmoments}
In this appendix, we derive Eq.~\eqref{eq:Z_n}. 
To this end, we need to compute the trace of the product of $\rho_{A,\alpha_j}$ which are given in Eq.~\eqref{eq:rho_A}. 
Applying the Baker-Campbell-Hausdorff formula and using the following commutation relation for arbitrary matrices $A$ and $B$, 
\begin{multline}
    \qty[ 
    \sum_{{\bf i,j}\in A\cap \Lambda_{\rm A}} 
    \mathbf{\Psi}_\mathbf{i}^\dag 
    A_{\mathbf{i,j}} 
    \mathbf{\Psi}_\mathbf{j},
    \sum_{{\bf i',j'}\in A\cap \Lambda_{\rm A}} 
    \mathbf{\Psi}_\mathbf{i'}^\dag 
    B_{\mathbf{i',j'}}
    \mathbf{\Psi}_\mathbf{j'}
    ]
    \\
    =
    \sum_{{\bf i,j}\in A\cap \Lambda_{\rm A}} 
    \mathbf{\Psi}_\mathbf{i}^\dag 
    [A,B]_{\mathbf{i,j}} 
    \mathbf{\Psi}_\mathbf{j}, 
\end{multline}
we can express the product of $\rho_{A,\alpha_j}$ as 
\begin{multline}
    \prod_{j=1}^{n} \rho_{A,\alpha_j} = 
    \det[\prod_{j=1}^n\frac{I+\Gamma_{\alpha_j}}{2}]
    \\
    \times\exp( \sum_{\mathbf{i},\mathbf{i'} \in A \cap \Lambda_A} \mathbf{\Psi}^{\dg}_{\mathbf{i}} K( \boldsymbol{\alpha} )_{\mathbf{i},\mathbf{i'}} \mathbf{\Psi}_{\mathbf{i'}}),
    \label{eq:product_density_matrix}
\end{multline}
where 
\begin{equation}
    K(\boldsymbol{\alpha}) = \log \left( \prod_{j=1}^{n} \frac{I - \Gamma_{\alpha_j}}{I + \Gamma_{\alpha_j}} \right).
    \label{eq:K(a)}
\end{equation}
Since the exponent on the right-hand side of Eq.~\eqref{eq:product_density_matrix} is quadratic in $\mathbf{\Psi}_\mathbf{i}$, it can be diagonalized as 
\begin{align}
    \prod_{j=1}^{n} \rho_{A,\alpha_j} = 
    \det(\prod_{j=1}^n\frac{I+\Gamma_{\alpha_j}}{2})
     e^{\sum_{i=1}^{2\ell L_y } \omega_i \gamma_i^\dag \gamma_i },
    \label{eq:product_density_matrix 2}
\end{align}
where $\omega_i$ is the $i$-th eigenvalue of $K(\boldsymbol{\alpha})$ and $\gamma_i~(\gamma_i^\dag)$ is a fermionic annihilation (creation) operator. 
Taking the trace of Eq.~\eqref{eq:product_density_matrix 2}, we obtain 
\begin{align}
    Z_n(\boldsymbol{\alpha},t)
    &= 
    \det[\prod_{j=1}^n\frac{I+\Gamma_{\alpha_j}}{2}]
    \prod_{j=1}^{2\ell L_y}
    (1+e^{\omega_i})
    \\
    &=
    \det[\prod_{j=1}^n\frac{I+\Gamma_{\alpha_j}}{2}]
    \det[I+e^{K(\boldsymbol{\alpha})}]. 
    \label{eq:product_density_matrix 3}
\end{align}
Substituting Eq.~\eqref{eq:K(a)} into Eq.~\eqref{eq:product_density_matrix 3}, we arrive at Eq.~\eqref{eq:Z_n}.

\section{Derivation of Eqs.~\eqref{eq:Z_nk/Z_nk}}
\label{app:n-thEA_GS}
Here, we derive Eq.~\eqref{eq:Z_nk/Z_nk} from Eq.~\eqref{eq:Z_nk}, which involves products and inverses of the block-Toeplitz matrices $I\pm \Gamma_{\alpha,k_y}$. As discussed in Sec.~\ref{sec:result_GS}, for $\ell \gg 1$, the multiple of the block-Toeplitz matrices can be approximated by the block-Toeplitz matrix generated by the product of the symbols of the factors. 
Namely, if we denote as $\mathrm{T}[\mathcal{G}]$ the $2\ell \times 2\ell$ block-Toeplitz matrix generated by the two-dimensional symbol $\mathcal{G}$, the following approximation holds for $\ell\gg1$, 
\begin{align}
    \mathrm{T}[\mathcal{G}] \mathrm{T}[\mathcal{G}']\simeq \mathrm{T}[\mathcal{GG'}]. \label{eq:TT}
\end{align}
Substituting $\mathcal{G}^{-1}$ into $\mathcal{G'}$ in the above equation, we obtain 
\begin{align}
\mathrm{T}[\mathcal{G}]^{-1}\simeq \mathrm{T}[\mathcal{G}^{-1}].\label{eq:Tinv} 
\end{align} 
Using Eqs.~\eqref{eq:TT} and \eqref{eq:Tinv}, the charged moments in Eq.~\eqref{eq:Z_nk} at $t=0$ can be approximated as 
\begin{align}
    \ln Z_{n,k_y}(\boldsymbol{\alpha},0)
    \simeq 
    \lim_{c\to1}
    \det \mathrm{T}[z_{n,\mathbf{k},c}(\boldsymbol{\alpha})], 
    \label{eq:lnZnk}
\end{align}
where 
\begin{multline}
    z_{n,\mathbf{k},c}(\boldsymbol{\alpha})
    =
    \prod_{j=1}^n \frac{I+c \,\mathcal{G}_{\mathbf{k},\alpha_j}(0)}{2}
    \\
    \times
    \qty[I+\prod_{j=1}^n \frac{I-c\,\mathcal{G}_{\mathbf{k},\alpha_j}(0)}{I+c\,\mathcal{G}_{\mathbf{k},\alpha_j}(0)}].
    \label{eq:z}
\end{multline}
Here, we introduced constant $c$ to avoid the singularity of $(I+\mathcal{G}_{\mathbf{k},\alpha_j})^{-1}$. 
Applying Widom-Szeg\H{o} theorem~\cite{widom1974asymptotic} to Eq.~\eqref{eq:lnZnk}, we obtain 
\begin{align}    
    \ln \frac{Z_{n,k_y}(\boldsymbol{\alpha},0)}{Z_{n,k_y}(\boldsymbol{0},0)} \simeq \lim_{c\to1}\ell \int_{-\pi}^{\pi} \frac{dk_x}{2\pi}  
    \ln \det z_{n,\mathbf{k},c}(\boldsymbol{\alpha}).
    \label{eq:chargedmoments_g_}
\end{align}
Here, we used $\det[z_{n,\mathbf{k},c}(\boldsymbol{0})]=1$. 
\par 
Using Eq.~\eqref{eq:z}, the integrand on the right-hand side of Eq.~\eqref{eq:chargedmoments_g_} can be written as 
\begin{multline}
    \ln\det z_{n,\mathbf{k},c}(\boldsymbol{\alpha})
    = \ln \prod_{j=1}^n \det\frac{I+c\,\mathcal{G}_{\mathbf{k},\alpha_j}(0)}{2}
    \\
    + 
    \ln \det[I+\prod_{j=1}^n \frac{I-c\,\mathcal{G}_{\mathbf{k},\alpha_j}(0)}{I+c\,\mathcal{G}_{\mathbf{k},\alpha_j}(0)}]. 
    \label{eq:chargedmoments_g}
\end{multline}
Since $I + c\,\mathcal{G}_{\mathbf{k},\alpha_j}$ has eigenvalues $1 \pm c$, the first term on the right-hand side of Eq.~\eqref{eq:chargedmoments_g} reduces to 
\begin{equation}
    \ln\prod_{j=1}^n\det\frac{I + c\,\mathcal{G}_{\mathbf{k},\alpha_j}(0)}{2}  = n\ln \left( \frac{1 - c^2}{4} \right).
    \label{eq:chargedmoments_firstterm}
\end{equation}
To evaluate the second term on the right-hand side of Eq.~\eqref{eq:chargedmoments_g}, we use the following formula for any $2\times2$ matrices $A$ and $B$,
\begin{equation}
    \det \left( A + B \right) = \det A + \det B + \Tr A \Tr B - \Tr AB. 
\end{equation}
Applying Eq.~\eqref{eq:chargedmoments_firstterm} and the above formula to the first and the second terms on the right-hand side of Eq.~\eqref{eq:chargedmoments_g}, respectively, we obtain 
\begin{multline}
    \ln \det  z_{n,\mathbf{k},c}(\boldsymbol{\alpha})
    =
    n \ln(\frac{1-c^2}{4})
    \\
    +
    \ln(2 
    + \Tr \prod_{j=1}^n \frac{I-c\,\mathcal{G}_{\mathbf{k},\alpha_j}(0)}{I + c\,\mathcal{G}_{\mathbf{k},\alpha_j}(0) }).
    \label{eq:chargedmoments_secondterm}
\end{multline}
Here, we used
\begin{equation}
    \det \left( \prod_{j=1}^n \frac{I - c\,\mathcal{G}_{\mathbf{k},\alpha_j}(0)}{I + c\,\mathcal{G}_{\mathbf{k},\alpha_j}(0)} \right) = 1, 
    \label{eq:second_det}
\end{equation}
which can be readily derived from the fact that $I+ c\,\mathcal{G}_{\mathbf{k},\alpha_j}$ has eigenvalues of $1 \pm c$. 
Using the identity 
\begin{equation}
    \left( I + c\,\mathcal{G}_{\mathbf{k},\alpha_j}(0) \right)^{-1} = \frac{I - c \,\mathcal{G}_{\mathbf{k},\alpha_j}(0)}{1-c^2},
\end{equation}
Eq.~\eqref{eq:chargedmoments_secondterm} reduces to 
\begin{multline}
    \ln \det z_{n,\mathbf{k},c}(\boldsymbol{\alpha})
    =
    n\ln(\frac{1-c^2}{4})
    \\
    +
    \ln(
    2+\frac{\Tr \prod_{j=1}^n (I - c\,\mathcal{G}_{\mathbf{k},\alpha_j}(0))^2}{(1-c^2)^{n}}
    ). 
    \label{eq:second_trace}
\end{multline}
Substituting it into Eq.~\eqref{eq:chargedmoments_g_} and taking the limit $c\to 1$, we obtain 
\begin{equation}
    \ln \frac{Z_{n,k_y}(\boldsymbol{\alpha},0)}{Z_{n,k_y}(\boldsymbol{0},0)} \simeq \ell \int_{-\pi}^{\pi} \frac{dk_x}{2\pi} \ln \Tr \prod_{j=1}^n \frac{I - \mathcal{G}_{\mathbf{k},\alpha_j}(0)}{2}.
    \label{eq:chargedmoments_Tr_I-g}
\end{equation}
Substituting Eq.~\eqref{eq:Gamma_symbol} into Eq.~\eqref{eq:chargedmoments_Tr_I-g} and using the cyclic property of trace, we obtain 
\begin{align}
    \ln \frac{Z_{n,k_y}(\boldsymbol{\alpha},0)}{Z_{n,k_y}(\boldsymbol{0},0)} \simeq  \ell \int_{-\pi}^{\pi} \frac{dk_x}{2\pi} \ln \Tr \prod_{j=1}^n \Pi_{\mathbf{k}}\sigma_x^{\alpha_j-\alpha_{j+1}}, 
    \label{eq:tr}
\end{align}    
where 
\begin{align}
    \Pi_{\mathbf{k}}=
    \frac{I-\sigma_x \sin \theta_\mathbf{k}
    -\sigma_z \cos \theta_\mathbf{k}}{2}.   
\end{align}
By simple algebra, one finds that $\Pi_{\mathbf{k}}\sigma_x^{\alpha_j-\alpha_{j+1}}\Pi_{\mathbf{k}}=\Pi_{\mathbf{k}}$ if $\alpha_j=\alpha_{j+1}$. 
This allows us to rewrite Eq.~\eqref{eq:tr} as 
\begin{align}
    \ln \frac{Z_{n,k_y}(\boldsymbol{\alpha},0)}{Z_{n,k_y}(\boldsymbol{0},0)} \simeq \ell \int_{-\pi}^{\pi} \!\frac{dk_x}{2\pi} \ln \Tr[(\Pi_{\mathbf{k}} \sigma_x)^{N^{(n)}(\boldsymbol{\alpha})}], 
    \label{eq:lnZn}
\end{align}
where $N^{(n)}(\boldsymbol{\alpha})=\sum_{j=1}^n |\alpha_j-\alpha_{j+1}|$ counts the number of domain walls, at which $\alpha_{j}\neq \alpha_{j+1}$, in the bit string $\boldsymbol{\alpha}\in \{0,1\}^n$.  
Using the fact that the matrix $\Pi_{\mathbf{k}} \sigma_x$ has the eigenvalues $-\sin \theta_\mathbf{k}$ and 0, we obtain $\Tr_{}[(\Pi_{\mathbf{k}} \sigma_x)^m]=(-\sin\theta_\mathbf{k})^m$. 
The right-hand side of Eq.~\eqref{eq:lnZn} can therefore be rewritten as 
\begin{align}
    \ln \frac{Z_{n,k_y}(\boldsymbol{\alpha},0)}{Z_{n,k_y}(\boldsymbol{0},0)} \simeq  \ell N^{(n)}(\boldsymbol{\alpha}) \int_{-\pi}^{\pi} \frac{dk_x}{4\pi} \ln \sin^2\theta_\mathbf{k}, 
\end{align}
which corresponds to Eq.~\eqref{eq:Z_nk/Z_nk} in the main text. 

\section{Derivation of Eq.~\eqref{eq:REA_t=0}}
\label{app:summation alpha}
Here, we give the derivation of Eq.~\eqref{eq:REA_t=0}. 
Substituting Eq.~\eqref{eq:Z/Z_n} into Eq.~\eqref{eq:REA_Z}, we have 
\begin{align}
    \Delta S_A^{(n)}(0)
    = 
    \frac{1}{1-n}
    \ln(
    \frac{1}{2^{n}}
    \sum_{\boldsymbol{\alpha}\in\{0,1\}^n}
    X_0^{N^{(n)}(\boldsymbol{\alpha})}). 
\end{align}
Recalling that $N^{(n)}(\boldsymbol{\alpha})=\sum_{j=1}^n |\alpha_j-\alpha_{j+1}|$ counts the number of domain walls between regions of consecutive 0's and 1's in the bit string $\boldsymbol{\alpha}=(\alpha_1,\alpha_2,...,\alpha_n)~(\alpha_i\in\qty{0,1})$, the summation over $\boldsymbol{\alpha}$ in the above equation can be rewritten as 
\begin{align}
    \Delta S_A^{(n)}(0)
    = 
    \frac{1}{1-n}
    \ln(
    \frac{1}{2^{n}}
    \sum_{m=0}^{\lfloor \frac{n}{2}\rfloor}
    \mqty(n\\2m)
    X_0^{2m}). 
    \label{eq:m}
\end{align}
Here, the binomial coefficient on the right-hand side represents the number of all possible bit strings $\boldsymbol{\alpha}$ that have $2m$ domain walls. 
Performing the summation over $m$, we obtain Eq.~\eqref{eq:REA_t=0}.

\bibliography{refs}       
\end{document}